\documentclass[a4paper,11pt]{article}

\pdfoutput=1
\usepackage{amsmath}
\usepackage{graphicx}

\usepackage[english]{babel}
\usepackage[utf8]{inputenc}
\usepackage{pdflscape}
\usepackage{enumerate}
\usepackage{amsbsy}
\usepackage{amsmath} %used by default
\usepackage{graphics}
\usepackage{mathrsfs}
\usepackage{wrapfig}
\usepackage{mathtools}

\usepackage{amsfonts}
\usepackage{pstricks}
\usepackage{color}
\usepackage{setspace}

\usepackage{jcappub_ver} % jcappub_ver load hyperref, load it last

\newcommand{\bea}{\begin{eqnarray}} \newcommand{\eea}{\end{eqnarray}}
\newcommand{\el}{\nonumber \\}
\newcommand{\re}[1]{(\ref{#1})}

\newcommand{\pat}{\partial}

\renewcommand{\sec}[1]{section \ref{#1}}

\newcommand{\para}{\paragraph}

\renewcommand{\a}{\alpha}
\renewcommand{\b}{\beta}
\renewcommand{\c}{\gamma}
\renewcommand{\d}{\delta}

\newcommand{\GN}{G_{\mathrm{N}}}
\newcommand{\ha}{\frac{1}{2}}

\newcommand{\rmd}{\mathrm{d}}

\newcommand{\ie}{i.e.\ }
\newcommand{\eg}{e.g.\ }

\newcommand{\av}[1]{\langle{#1}\rangle}
\newcommand{\Mpl}{M_{{}_{\mathrm{Pl}}}}

\newcommand{\lc}{\lambda_\text{c}}
\newcommand{\thetat}{\tilde{\theta}}
\newcommand{\sigmat}{\tilde{\sigma}}
\newcommand{\htt}{\tilde{h}}

\title{The effect of dark matter discreteness on light propagation}

\author[a, b]{Sofie Marie Koksbang}
\author[b]{and Syksy R\"{a}s\"{a}nen}

\affiliation[a]{CP$^3$-Origins, University of Southern Denmark, Campusvej 55, DK-5230 Odense M, Denmark}

\affiliation[b]{University of Helsinki, Department of Physics and Helsinki Institute of Physics \\ P.O. Box 64, FIN-00014 University of Helsinki, Finland}

\emailAdd{koksbang@cp3.sdu.dk}
\emailAdd{syksy.rasanen@iki.fi}

\abstract{
Light propagation in cosmology is usually studied in the geometrical optics approximation which requires the spacetime curvature to be much smaller than the light wavenumber. However, for non-fuzzy particle dark matter the curvature is concentrated in widely separated spikes at particle location. If the particle mass is localised within a Compton wavelength, then for masses $\gtrsim10^4$ GeV the curvature is larger than the energy of CMB photons.

We consider a post-geometrical optics approximation that includes curvature. Photons gain a gravity-induced mass when travelling through dark matter, and light paths are not null nor geodesic. We find that the correction to the redshift is negligible. For the angular diameter distance, we show how the small average density emerges from the large local spikes when integrating along the light ray. We find that there can be a large correction to the angular diameter distance even for photon energies much larger than the curvature. This may allow to set a strong limit on the mass of dark matter particles. We discuss open issues related to the validity of our approximations.
}

\begin{document}

\begin{flushleft}
	\hfill		 HIP-2021-25/TH \\
\end{flushleft}
 
\setcounter{tocdepth}{3}

\setcounter{secnumdepth}{3}

\maketitle

\section{Introduction} \label{sec:intro}

\para{Beyond geometrical optics.}

Cosmological observations and hence our conclusions about the universe at large are mainly based on light (although neutrinos, charged particles and, most recently, gravitational waves are also observed). The analysis of light propagation is usually based on the geometrical optics approximation \cite{PhoneBook} (page 570), \cite{Lensing} (page 93). It assumes that the wavelength is much smaller than the other scales in the problem, namely the length over which the amplitude and polarisation change, and the spacetime curvature radius. In this case light propagation can be described in terms of a local plane wave orthogonal to a null geodesic ray. This is a version of the WKB approximation and is also called the eikonal approximation.

The geometrical optics approximation is known to fail in gravitational lensing when the lens size is comparable to the wavelength, as the wavefront is not constant over a few wavelengths. Instead, the wave-optics approximation is used (see \eg \cite{WaveOptics_4, WaveOptics_5, WaveOptics_2,WaveOptics_3, Turyshev:2021eoh} and references therein\footnote{We focus on light; for discussion of geometrical optics and wave optics for gravitational waves, see \cite{GW_1,GW_2,GW_3, WaveOptics_GW, GW_4,GW_5}.}). A different issue is related to spacetime curvature. In general, vacuum solutions of the Maxwell equation in curved spacetime do not follow null geodesics, as is well known \cite{DeWitt:1960fc, MaxwellEFE_1,MaxwellEFE_4, Copi:2020qur, critique_1}. Expanding in the short wavelength approximation beyond the leading terms \cite{HigherOrder_1, Dolan:2018ydp}, the covariant derivatives in the equation of motion lead to spin-dependent light propagation, known as the gravitational Hall effect \cite{Oancea:2019pgm, Oancea:2020khc, Frolov:2020uhn}. However, the Maxwell equation in curved spacetime also involves the Ricci tensor, and the related curvature radius $R$ can be much smaller than the wavelength.

If dark matter consists of well-separated particles with mass $m$, approximated as spheres of constant density with radius equal to the Compton wavelength $\lc=2\pi/m$, we have $R^{-2}\sim\rho/\Mpl^2\sim 10^{-3} m^4/\Mpl^2$, where $\Mpl$ is the Planck mass and $\rho$ is the local energy density. The lowest energy of observed cosmic photons is determined by the fact that the ionosphere is opaque to light frequencies below $\sim$ 10 MHz, corresponding to light wavelength $\sim10$ m and energy $E\sim10^{-8}$ eV. (None of the radio telescopes in space have been sensitive to such small frequencies.) Setting $\rho/\Mpl^2=E^2$ gives $m\sim10$ GeV. If dark matter is heavier than this, the curvature is larger than the smallest observed photon energy, so curvature effects may be relevant for some electromagnetic observations. By this estimate, the effect of electrons is negligible. The density of nuclei is $\rho\sim10^{-3}$ GeV$^{4}$, which gives $\rho/(E^2 \Mpl^2)\sim10^{-6}$, also negligible. Also, light waves cannot pass through charged particles in the same way as through dark matter. The highest energy photons observed have $E\sim10^5$ GeV, and the corresponding dark matter mass is $m\gtrsim10^{12}$ GeV. Cosmic microwave background (CMB) photons at redshift $z$ have energy $\sim10^{-3}(1+z)$ eV, so for $m\gtrsim10^4\sqrt{1+z}$ GeV, the curvature radius at dark matter particle position is smaller than the photon wavelength. The cosmological average number density of dark matter particles is $\sim (1+z)^3 (m/\text{GeV})^{-1}$m$^{-3}$, so for all these masses the distance between dark matter particles is many orders of magnitude larger than their size, even when local clustering is taken into account. If the curvature radius is smaller than the photon wavelength, the photon travels through an environment characterised by well-separated, narrow and high curvature spikes, rather than a smooth background as assumed in geometrical optics. We will see that the naive estimate above based on the curvature radius is too conservative, as the angular diameter distance involves not just the density but also its derivative.

While the effect of clumping on light propagation beyond perturbation theory has been considered before, including with discrete distributions of matter \cite{Zeldovich:1964, Dashevskii:1965, Dashevskii:1966, Bertotti:1966, Gunn:1967, Dyer:1974, Dyer:1981, Kasai:1990hd, Kibble:2004tm, Clifton:2009jw, Clifton:2009bp, Clifton:2010fr, Clarkson:2011br, Clifton:2011mt, Bruneton:2012cg, Bruneton:2012ru, Larena:2012vn, Liu:2015bya, Sanghai:2015wia, Fleury:2015rwa, Bentivegna:2016fls, Sanghai:2017yyn, Bentivegna:2018koh, Fleury:2017owg, Fleury:2018cro, Fleury:2018odh}, these studies have been in the context of the geometrical optics approximation, and have also considered the possible effect of inhomogeneities on the average expansion rate \cite{Buchert:2011sx}. We are concerned only with light propagation, and start from the Maxwell equation. In \sec{sec:post-geometric} we derive a post-geometrical approximation that includes curvature and calculate the effect of dark matter clumpiness on redshift and angular diameter distance. We summarise our findings and highlight open questions in \sec{sec:conclusion}. In appendix \ref{sec:area} we calculate the relation between the angular diameter distance and light bundle area expansion rate, and in appendix \ref{sec:duality} we show that distance duality holds in our post-geometrical approximation.

\section{Post-geometrical optics} \label{sec:post-geometric}

\subsection{Post-geometrical approximation}

\para{The equation of motion.}

We start from the action $S=-\int\rmd^4 x\sqrt{-g}\frac{1}{4}F_{\a\b} F^{\a\b}$ in curved spacetime, where the electromagnetic field strength $F_{\a\b}$ is written in terms of the vector potential $A^\a$ as $F_{\a\b}=\pat_\a A_\b-\pat_\b A_\a$. Varying the action with respect to $A^\a$ gives the equation of motion (we consider an uncharged medium):
\begin{align} \label{eom}
  \Box A^{\a} - \nabla^\a \nabla_\b A^\b - R^\a{}_\b A^\b = 0\ ,
\end{align}
where $\Box\equiv g^{\a\b}\nabla_\a \nabla_\b$ and $R_{\a\b}$ is the Ricci tensor. We take the Lorenz gauge condition
\bea \label{Lorenz}
  \nabla_\a A^\a=0 \ ,
\eea
so \re{eom} becomes a curved spacetime generalisation of the wave equation, which depends explicitly on the Ricci tensor.

As in the geometrical optics approximation, we consider the local plane wave form \cite{PhoneBook} (page 570), \cite{Lensing} (page 93)
\begin{align}\label{eq:A_expansion}
	A^{\a} = \sum_{n=0}^\infty \text{Re} ( A^\a_n \epsilon^n e^{i S/\epsilon} ) \ ,
\end{align}
where the amplitude is expanded in a series of the constant parameter $\epsilon\ll1$, which represents the ratio of the wavelength to other relevant scales that are considered large. Unlike in the geometrical optics approximation, we do not assume that the spacetime curvature radius is large compared to the wavelength.

Inserting the expansion \re{eq:A_expansion} into the equation of motion \re{eom} gives, in the Lorenz gauge,
\bea \label{eomexp}
    \!\!\!\!\!\!\!\!\! 0 &=& \sum_{n=0}^\infty \text{Re} \left[ \epsilon^n \left( - \epsilon^{-2} k^2 A^\a_n + i \epsilon^{-1} \nabla_\b k^\b A^\a_n + 2 i \epsilon^{-1} k^\b \nabla_\b A^\a_n + \Box A^\a_n - R^\a{}_\b A^\b_n \right) e^{i S/\epsilon} \right] \ ,
\eea
where $k_\a\equiv\pat_\a S$ and $k^2\equiv k_\a k^\a$. In the geometrical optics approximation, the derivatives of $A^\a_n$ are taken to be of the same order of smallness as $A^\a_n$, meaning that amplitude and polarisation change slowly compared to the wavelength. We make the same assumption. In the geometrical optics approximation, $R^\a{}_\b$ is taken to be independent of $\epsilon$. However, if the curvature is large, it will instead be the dominant term in the equation of motion, and will set the scale of the derivatives. We correspondingly assume that $R^\a{}_\b$ is of order $\epsilon^{-2}$. We can still formally smoothly take the limit $R^\a{}_\b\to0$ to recover geometrical optics. However, in between, $R^\a{}_\b$ would contribute at order $\epsilon^{-1}$, so the approximation may not cover regimes of intermediate curvature properly. Expanding the Lorenz gauge condition \re{Lorenz} in the same way, we have
\bea \label{Lorenzexp}
    0 &=& \sum_{n=0}^\infty \text{Re} \left[ \epsilon^n \left( \epsilon^{-1} i k_\a A^\a_n + \nabla_\a A^\a_n \right) e^{i S/\epsilon} \right] \ .
\eea

Considering the equation of motion \re{eomexp} at orders $\epsilon^{-2}$ and $\epsilon^{-1}$ gives
\bea
  \label{go1} R_{\a\b} a^\b + k^2 a_\a &=& 0 \\
  \label{go2} R_{\a\b} b^\b + k^2 b_\a &=& a_\a \nabla_\b k^\b + 2 k^\b \nabla_\b a_\a \ ,
\eea
where we have denoted $A^\a_0\equiv a^\a$, $A^\a_1\equiv i b^\a$. Expanding the Lorenz gauge condition \re{Lorenzexp} to orders $\epsilon^{-1}$ and $\epsilon^0$ gives (for any vectors $v^\a$ and $w^\a$, we denote $v\cdot w\equiv g_{\a\b} v^\a w^\b$)
\bea \label{Lorenz2}
  \label{go3} a\cdot k &=& 0 \\
  \label{go4} b \cdot k &=& \nabla_\a a^\a \ .
\eea
This is the same result as in geometrical optics and shows that the light ray is transverse. Contracting \re{go2} with $a^\a$ and using \re{go1}, we obtain
\bea \label{pn}
  0 &=& \nabla_\b ( a \cdot a \, k^\b ) \ ,
\eea
so photon number is conserved, as in geometrical optics. In geometrical optics $k^\a$ is null, so it follows from the transversality condition \re{go3} that $a^\a$ is spacelike or null. Here that is not the case, but we assume, as in geometrical optics, that $a^\a$ is spacelike and decompose it as $a^\a=a f^\a$, where $a>0$ and $f\cdot f=1$. Using this decomposition in \re{go2} and applying \re{pn}, we find
\bea \label{pol}
  k^\b \nabla_\b f_\a &=& \frac{1}{2 a} ( R_{\a\b} b^\b + k^2 b_\a ) \ .
\eea
This shows that the polarisation vector is not necessarily parallel transported along the light ray. Using the decomposition of $a^\a$ in \re{go1} and contracting with $f^\b$, we obtain
\bea \label{null}
  k^2 &=& - R_{\a\b} f^\a f^\b \ ,
\eea
so in general the light rays are not null. Taking a covariant derivative of \re{null} and using the fact that $\nabla_\b k_\a=\nabla_\a k_\b$ (since $k_\a=\pat_\a S$) we get
\bea \label{knk}
  k^\b \nabla_\b k_\a &=& \ha \nabla_\a k^2 = - \ha \nabla_\a ( R_{\a\b} f^\a f^\b ) \ ,
\eea
showing that in general the light rays are not geodesic.

\para{The energy-momentum tensor.}

Consider a unit timelike vector $u^\a$ ($u\cdot u=-1$). Without loss of generality, we can decompose the energy-momentum tensor as
\bea \label{Tab}
  T_{\a\b} = ( \rho + p ) u_\a u_\b + p g_{\a\b} + 2 q_{(\a} u_{\b)} + \Pi_{\a\b} \ ,
\eea
where $\rho$ is the energy density, $p$ is the pressure, $q_\a$ is the energy flux, and $\Pi_{\a\b}$ is the anisotropic stress. Both $q_\a$ and $\Pi_{\a\b}$ are orthogonal to $u^\a$, and $\Pi_{\a\b}$ is traceless and symmetric. From the Einstein equation $R_{\a\b}-\ha R g_{\a\b}=\Mpl^{-2} T_{\a\b}$, where $R\equiv g^{\a\b} R_{\a\b}$, we get
\bea \label{Rab}
  \Mpl^2 R_{\a\b} &=& ( \rho + p  ) u_\a u_\b + \ha ( \rho - p ) g_{\a\b} + 2 q_{(\a} u_{\b)} + \Pi_{\a\b} \ .
\eea
Inserting \re{Rab} into \re{go1} and using \re{null}, we obtain
\bea \label{fu}
  [ ( \rho + p ) u \cdot f + q \cdot f ] u_\a &=& [ ( \rho + p )  (u\cdot f)^2 + 2 u \cdot f q \cdot f  + \Pi_{\c\d} f^\c f^\d ] f_\a \el
  && - u \cdot f q_\a - \Pi_{\a\b} f^\b \ .
\eea
As $u_\a$ is timelike and the right-hand side is a spacelike vector, their coefficients have to vanish separately. From the left-hand side we get $( \rho + p ) u \cdot f + q \cdot f=0$. As there is no reason for the directions of the energy flux and the polarisation to be locally related, the most reasonable way to satisfy this condition is $u\cdot f=0$, $q_\a=0$. The right-hand side condition then reduces to $\Pi^\a{}_\b f^\b=\Pi_{\c\d} f^\c f^\d f^\a$. By the same argument, the most reasonable solution is $\Pi_{\a\b}=0$. These constraints are a limitation of our post-geometrical approximation. When the curvature (and hence the energy-momentum tensor) is directly coupled to light propagation, the local plane wave approximation is not valid if the energy-momentum tensor introduces new privileged spatial directions, because in the leading order plane wave approximation only the propagation direction given by $k^\a$ and the polarisation direction $f^\a$ are relevant. (Although $u^\a$ is timelike, the projection of $f^\a$ orthogonal to it introduces a new spatial direction unless $u\cdot f=0$.) With $u\cdot f=0$, $q_\a=0$, $\Pi_{\a\b}=0$, \re{null} reduces to
\bea \label{null2}
  k^2 &=& - \frac{\rho - p}{2 \Mpl^2} \ .
\eea
For matter satisfying $\rho\geq p$, the tangent vector $k^\a$ is timelike or null, so the gravitational effect of matter slows down light, analogously to refraction when passing through charged matter.

Applying \re{Tab} to the polarisation propagation equation \re{pol} gives
\bea \label{pol2}
  k^\b \nabla_\b f_\a &=& \frac{1}{2 \Mpl^2 a} ( \rho + p ) u \cdot b \, u_\a \ .
\eea
The right-hand side is in general non-zero, so the polarisation vector rotates, \ie there is gravitational circular birefringence, unlike in geometrical optics. This is required by the condition $u\cdot f=0$, because $u^\a$ is in general not parallel transported along the light ray. The magnitude of the birefringence can be calculated by finding $u\cdot b$ from the requirement that the condition $u\cdot f=0$ is preserved along the light ray, although we should then check overall consistency to next order in $\epsilon$.

The propagation equation \re{knk} for $k^\a$ reduces to the simple expression
\bea \label{knk2}
  k^\b \nabla_\b k_\a &=& - \frac{1}{4 \Mpl^2} \pat_\a ( \rho - p ) \ .
\eea
The results \re{null2} and \re{knk2} allow us to calculate the post-geometrical correction to redshift and angular diameter distance.

\subsection{Redshift} \label{sec:redshift}

\subsubsection{Expression for the redshift}

\para{Dispersion relation.}

We decompose the light ray tangent vector as
\bea \label{kdec}
  k^\a &=& E ( u^\a + v e^\a ) \ ,
\eea
where $E=-u \cdot k$, $u \cdot e=0$, $e \cdot e = 1$, and $v\equiv\sqrt{1+k^2/E^2}=\sqrt{1-\frac{\rho-p}{2 \Mpl^2 E^2}}$. Correspondingly, the dispersion relation is
\bea
  E^2 = \vec k^2 + \frac{\rho - p}{2 \Mpl^2} \ ,
\eea
where $|\vec k|=v E$ is the spatial wavenumber, and as usual the wavelength is $2\pi/|\vec k|$. Light acquires the gravitationally induced mass $M\equiv\sqrt{(\rho-p)/(2\Mpl^2)}$, of the order of the curvature scale. For a light ray passing from vacuum to matter the wavenumber has to satisfy $E=|\vec k|>M$, otherwise the wave cannot enter, and the oscillating solution turns into an exponentially decaying mode. This is analogous to the phenomenon that light cannot propagate in materials whose plasma frequency exceeds the light frequency. The refractive index is $n=v^{-1}=(1-M^2/E^2)^{-1/2}$. In a region where $M\neq0$, we can write $k^\a=M w^\a$, where $w^\a$ is a unit timelike vector, and the propagation equation \re{knk} reads $w^\b\nabla_\b w_\a=-M^{-1}\hat\nabla_\a M$, where $\hat\nabla_\a$ is the derivative projected orthogonally to $w^\a$. This can be compared to the equation for an observer comoving with an ideal fluid, $w^\b\nabla_\b w_\a=-(\rho+p)^{-1}\hat\nabla_\a p$.

\para{Redshift.}

The redshift is given by
\begin{align}
	1+z = \frac{E_\text{s}}{E_\text{o}} \ ,
\end{align}
where s refers to source and o to observation. To find the redshift, we proceed as in geometrical optics (see \eg \cite{light_stat1, light_stat2}), and take a derivative with respect to the affine parameter $\lambda$ along the light ray,
\bea \label{E1}
  \frac{\rmd E}{\rmd\lambda} &=& -k^{\b}\nabla_{\b}\left(u^{\a}k_\a \right) 
  % \el &=& - k^\a k^\b \nabla_\b u_\a - u^\a k^\b \nabla_\b k_\a \el &=&
  = - k^\a k^\b \nabla_\b u_\a - \ha u^\a \nabla_\a k^2 \ ,
\eea
where we have used \re{knk}. We decompose $\nabla_\b u_\a$ as usual (for overviews of the covariant formalism, see \cite{Ellis:1971pg, Tsagas:2007yx})
\bea \label{udec}
  \nabla_\b u_\a &=& \frac{1}{3} \theta h_{\a\b} + \sigma_{\a\b} + \omega_{\a\b} -\dot u_\a u_\b \ ,
\eea
where $h_{\a\b}\equiv g_{\a\b}+u_\a u_\b$, $\theta\equiv\nabla_\a u^\a$ is the volume expansion rate, $\sigma_{\a\b}$ is the traceless symmetric shear tensor, $\omega_{\a\b}$ is the antisymmetric vorticity tensor, $\dot u^\a\equiv u^{\b}\nabla_{\b}u^{\a}$ is the acceleration vector, and dot denotes $u^\a \nabla_\a$. The tensors $\sigma_{\a\b}$ and $\omega_{\a\b}$ and the vector $\dot u^\a$ are orthogonal to $u^\a$. Inserting \re{kdec} and \re{udec} into \re{E1}, we get
\begin{align} \label{dE}
	\frac{1}{E^2}\frac{\rmd E}{\rmd\lambda} = - \frac{1}{3} v^2 \theta - v ^2 \sigma_{\a\beta} e^{\a}e^{\b} -v e \cdot \dot u - \frac{1}{2 E^2} (k^2)\dot{} \ .
\end{align}
Integrating, we obtain
\bea \label{zgen}
	\ln\left(1+z\right) &=& \int_{\lambda_\text{s}}^{\lambda_\text{o}} \rmd\lambda E^{-1} \left[ \frac{1}{3} v^2 \theta + v ^2 \sigma_{\a\beta} e^{\a}e^{\b} + v e \cdot \dot u + \frac{1}{2 E^2}(k^2)\dot{} \right] \el
	&=& \int_{\lambda_\text{s}}^{\lambda_\text{o}} \rmd\lambda E^{-1} \left[ v^2 \left( \frac{1}{3} \theta + \sigma_{\a\beta} e^{\a}e^{\b} \right) - \frac{v}{\rho+p} e^\a \pat_\a p - \frac{\dot\rho - \dot p}{4 \Mpl^2 E^2} \right] \ ,
\eea
where in the second equality we have used \re{null2} and the spatial part of the continuity equation,
\bea \label{udot}
  \dot u_\a = -\frac{1}{\rho+p} h_\a{}^\b\pat_\b p \ .
\eea
The result \re{zgen} reduces to the geometrical optics expression when $k^2=0$, \ie $v=1$. In the opposite limit $|\vec k|=0$, \ie $v=0$ and $E^2=M^2=(\rho-p)/(2\Mpl^2)$, the first two terms vanish as the curvature-induced mass becomes so large that light cannot propagate inside the matter.

\subsubsection{Size of the curvature correction}

\para{Matter model.}
 
Dark matter energy density consists of widely separated sharp spikes, which we model as
\begin{align}
	\rho = m \sum_n W(r_n) \ ,
\end{align}
where $n$ labels particles, $W$ is a window function that gives the density distribution of a particle (related to its wavefunction), and $r_n\equiv|\vec r-\vec r_n|$ is the proper distance to the centre of particle $n$, located at $\vec r_n$, along a line orthogonal to $u^\a$. We take $W$ to be a Gaussian of width equal to the Compton wavelength $\lc=2\pi/m$, so
\bea \label{rho}
  \rho &=& m \sum_n \frac{1}{(2\pi)^{3/2} \lc^3} e^{-\frac{r_n^2}{2\lc^2}} = \sqrt{\frac{2}{9\pi}} \rho_0 \sum_n e^{-\frac{r_n^2}{2\lc^2}} \ ,
\eea
where $\rho_0\equiv m/(\frac{4}{3}\pi \lc^3)$ is the energy density averaged over one particle. The distance between dark matter particles is so large that overlap is negligible. We model the region inside the particle using general relativity, so although the pressure is negligible, we need $p\neq0$ for the particle not to collapse, and the pressure gradient may be important. We assume $m\ll\Mpl$, so the curvature is much smaller than the Planck scale, and we treat spacetime near and inside each particle as perturbed Minkowski space, mostly working to leading order (which often means neglecting the perturbations). We do not consider non-gravitational interactions between dark matter and light.

\para{Correction to redshift.}

We have $\dot\rho=0$ everywhere. This is not in contradiction with the expansion of space and decrease of average density. The time part of the continuity equation, $\dot\rho+\theta\rho=0$, is satisfied because inside particles the expansion rate is zero and outside the density is zero. (The separation is of course blurry as the particles do not have sharp boundaries.) In the redshift integral \re{zgen} the factor $v^2$ in front of $\theta$ is essentially unity whenever $\theta$ is non-zero. The shear is zero inside the particles, because the solution is spherically symmetric and static. The acceleration term integrates to zero, because the spacetime is static inside the particle, so the light gains as much energy going in as it loses coming out. The last term is zero because $\dot\rho=\dot p=0$. There is in principle a correction to the redshift from the missing expansion and slower light travel inside the particles, but this is negligible because the particles occupy only a tiny fraction of the volume. We consider only observers whose time direction is orthogonal to the polarisation, but the redshift of different observers is related to each other by a Lorentz boost as usual. So the curvature correction to the redshift is negligible. We now turn to the angular diameter distance, where the situation is different.

\subsection{Angular diameter distance} \label{sec:distance}

\subsubsection{Expression for the angular diameter distance} \label{sec:DA}

\para{Angular diameter distance and the light bundle.}

In order to determine the angular diameter distance $D_A$, we have to find the area expansion rate of a bundle of light rays orthogonal to $k^\a$ and to a four-velocity. We present the details in appendix \ref{sec:area}. In our post-geometrical approximation (unlike in geometrical optics), the area expansion rate depends on the choice of four-velocity. However, when calculating $D_A$, the velocity-dependent part is subdominant and can be neglected.

It is useful to introduce a tensor that projects orthogonally to $k^\a$ and $u^\a$,
\bea \label{htt}
  \htt_{\a\b} &\equiv& g_{\a\b} + u_\a u_\b - e_\a e_\b \el
  &=& g_{\a\b} + \frac{k^2}{E^2+k^2} u_\a u_\b + \frac{2 E}{E^2+k^2} u_{(\a} k_{\b)} - \frac{1}{E^2+k^2} k_\a k_\b \ .
\eea
We decompose $\nabla_\b k_\a$ into parts projected in the direction of and orthogonal to $\htt_{\a\b}$ as
\bea \label{nablakdec}
  \nabla_\b k_\a = \frac{1}{2} \thetat \htt_{\a\b} + \sigmat_{\a\b} + P_{\a\b} \ ,
\eea
where the area expansion rate is $\thetat\equiv \htt^{\a\b} \nabla_\b k_\a$, the light shear is $\sigmat_{\a\b}=\htt_\a{}^\c \htt_\b{}^\d \nabla_\d k_\c-\ha \htt_{\a\b} \thetat$, and $P_{\a\b}\equiv (\d_\a{}^\c \d_\b{}^\d - \htt_\a{}^\c \htt_\b{}^\d ) \nabla_\d k_\c$. Unlike in geometrical optics, $P_{\a\b}$ is neither traceless nor orthogonal to $k^\a$. As shown in appendix \ref{sec:area} we have $D_A\propto\exp\left( \ha\int\rmd\lambda \nabla_\a k^\a \right)=\exp\left[ \ha\int\rmd\lambda ( \thetat + P^\a{}_\a ) \right]$.

\para{Integrating along the light ray.}

In order to integrate $\thetat$ over the light ray, we do the same as in geometrical optics, namely take the derivative, and then decompose and integrate it. We have
\bea \label{dtheta}
  \frac{\rmd\thetat}{\rmd\lambda} &=& k^{\b}\nabla_{\b} \nabla_\a k^{\a} - \frac{\rmd P^\a{}_\a}{\rmd\lambda} \el
  &=& - \frac{1}{2}\tilde{\theta}^2 - 2\sigmat^2 - R_{\a\b} k^{\a}k^{\b} + \frac{1}{2}\Box k^2 - P_{\a\b} P^{\a\b} - \frac{\rmd P^\a{}_\a}{\rmd\lambda} \ ,
\eea
where $\sigmat^2\equiv\ha\sigmat_{\a\b}\sigmat^{\a\b}$, and we have applied the definition of the Riemann tensor in terms of a commutator of covariant derivatives. In the geometrical optics limit, only the first three terms remain. Expanding $P_{\a\b} P^{\a\b}$ using \re{htt}, we get
\bea \label{AA}
  P_{\a\b} P^{\a\b} &=& \frac{1}{v^2 E^2} \left\{ \frac{k^2-E^2}{2 v^2 E^2} [(k^2)\dot{}]^2 + \frac{1}{v^2 E} (k^2)\dot{} (k^2)' - \frac{1}{4 v^2 E^2} [(k^2)']^2 - 2 k^2 \dot k_\a \dot k^\a \right. \el
  && \left. - 4 E k_\a' \dot k^\a + 2 k_\a' k^\a{}' - \frac{\dot k \cdot u}{v^2 E} \left[ 2 k^2 (k^2)\dot{} + E (k^2)' \right] - \frac{(k^2)^2}{v^2 E^2} ( \dot k \cdot u )^2 \right\} \el
  &=& \frac{1}{v^2 E^2} \left\{ - \frac{1}{16 \Mpl^4} [ e^\a\pat_\a (\rho-p)]^2 + \frac{\rho}{\Mpl^2} \dot k_\a \dot k^\a + \frac{E}{\Mpl^2} \dot k^\a \pat_\a (\rho-p) \right. \el
  && + \frac{1}{8 \Mpl^4} \pat_\a (\rho-p) \pat^\a (\rho-p) + \frac{E^2}{2 \Mpl^2\rho} e^\a\pat_\a p \, e^\b \pat_\b ( \rho - p ) \el
  && \left. - \frac{1}{4 \Mpl^4} ( e^\a \pat_\a p )^2 \right\} \ ,
\eea
where prime denotes $k^\a\nabla_\a$, and in the second equality we have used \re{null2}, \re{knk2} and \re{udot}, and taken into account $\dot\rho=\dot p=0$, $p\ll\rho$, and that inside the particles $\dot E=0$. Spatial derivatives come from contracting $\nabla_\b k_\a$ with $k^\a$ (entering from the definition \re{htt} of $\htt_{\a\b}$), while both $u^\a$ and $k^\a$ contribute to the time derivatives. In geometrical optics, in a spatially flat Friedmann--Lema\^itre--Robertson--Walker (FLRW) universe we have $\nabla_\a k^\a=2E(H+L^{-1})$, where $H$ is the Hubble parameter and $L$ is the proper spatial distance to the source. Because the density changes on a small length scale, the spatial derivative terms are large compared to the geometrical optics terms. However, it is not immediately obvious whether the post-geometrical correction to the angular diameter distance is large, because these terms are non-zero only in a small fraction of the volume.

Let us first show that we can neglect $\int\rmd\lambda P^\a{}_\a$, and hence the difference between $\nabla_\a k^\a$ and $\thetat$ inside the integral. Using \re{htt}, we have
\bea \label{P}
  \int\rmd\lambda P^\a{}_\a &=& - \int\rmd\lambda \left\{ \frac{k^2}{E^2+k^2} u \cdot \dot k + \frac{1}{E^2 + k^2}\left[ E (k^2)\dot{} - \ha (k^2)' \right] \right\} \el
  &=& - \int\rmd\lambda \left\{ \frac{1}{2 v^2 \Mpl^2 E^2}\left[ \dot u \cdot k \rho + \ha E e^\a \pat_\a (\rho-p) \right] \right\} \el
  &=& - \int\rmd\lambda \frac{1}{4 v \Mpl^2 E} e^\a \pat_\a ( \rho - 3 p ) \ ,
\eea
where on the second line we have used \re{null2}, taking into account that $\dot\rho=\dot p=0$, $p\ll\rho$, and that inside the particles $\dot E=0$ (as the metric is static, photon energy depends on position, but not explicitly on time), and on the third line we have used \re{udot}. As $\rho$ and $p$ inside each particle depend only on the distance from the centre, the directional derivatives become sums over dot products. For each particle $n$, we have $e^\a \pat_\a = \d_{ij} e^i (r^j-r_n^j)/r_n \pat_{r_n}$. The dot product gives the cosine of the angle between the light ray and a line from the centre of particle $n$. When we average over many rays within a tube of area $A$, we integrate over the entry angle, with a distribution that is flat in the cosine of the entry angle. (Summing over a large number of particles for a single ray would give the same distribution.) Assuming the change in light ray direction is small or symmetric around the centre of each particle, we get zero.

As the contribution of $P^\a{}_\a$ is negligible, we have from \re{dtheta} and \re{AA}
\bea \label{thetaint}
  \int\rmd\lambda \nabla_\a k^\a &=& \int\rmd\lambda \thetat \el
  &=& \int\rmd\lambda \int\rmd\lambda' \left( - \frac{1}{2} \thetat^2 - 2 \sigmat^2 -R_{\a\b} k^{\a}k^{\b} + \frac{1}{2}\Box k^2 -P_{\a\b} P^{\a\b} \right) \el
  &=& \int\rmd\lambda \int\rmd\lambda' \Bigg\{ - \frac{1}{2} \thetat^2 - 2 \sigmat^2 \underbrace{ - E^2 \frac{\rho + p}{\Mpl^2} \left[ 1 - \frac{(\rho-p)^2}{4\Mpl^2 E^2 (\rho+p)} \right]}_{\text{term 1}} \el
  && \underbrace{- \frac{1}{4 \Mpl^2} \Box ( \rho - p )}_{\text{term 2}} \underbrace{- P_{\a\b} P^{\a\b}}_{\text{term 3}} \Bigg\} \ ,
\eea
where we have used \re{Rab} and \re{null2}. The first two terms are the same as in geometrical optics. Let us consider the three post-geometrical correction terms in turn.

\subsubsection{Size of the curvature correction} \label{sec:corr}

\para{Correction term 1.}

Over one wavelength, a light ray typically passes through no particles (except for long light wavelength and small dark matter mass). However, measurements are made with finite beam size, and the effect of dark matter is smeared over the width of the beam. Let us consider an interval $\Delta\lambda$ that corresponds to a time interval $\Delta t$ that is small compared to the timescale of cosmological evolution but large compared to the distance between dark matter particles, and average over the beam area $A$, approximating spacetime in this tube as perturbed Minkowski space. The first post-geometrical correction term in \re{thetaint} involves the integral
\bea \label{rho2}
  \int_{\Delta\lambda} \rmd\lambda E^2 \frac{\rho}{\Mpl^2} \left( 1 - \frac{\rho}{4\Mpl^2 E^2} \right) &=& \int_{\Delta t}\rmd t E \frac{\rho}{\Mpl^2} \left(1 - \frac{\rho}{4 \Mpl^2 E^2} \right) \el
  &=& \frac{1}{A} \int_{V} \rmd^3 r E \frac{\rho}{\Mpl^2} \left(1 - \frac{\rho}{4 \Mpl^2 E^2} \right) \el
  &=& \frac{1}{A} \sum_n \int_{V} \rmd^3 r E \frac{m}{\Mpl^2} W(r_n) \left[ 1 - \frac{m}{4\Mpl^2 E^2} W(r_n) \right] \el
  &=& \frac{\Delta t}{V} E N \frac{m}{\Mpl^2} \left( 1 - \frac{\rho_0}{24 \sqrt{\pi} \Mpl^2 E^2} \right) \el
  &=& \int_{\Delta\lambda} \rmd\lambda E^2 \frac{\av{\rho}}{\Mpl^2} \left( 1 - \frac{\rho_0}{24 \sqrt{\pi} \Mpl^2 E^2} \right) \ ,
\eea
where we have used $\rmd\lambda E = \rmd t$, taken into account $p\ll\rho$, inserted the Gaussian density distribution \re{rho}, and neglected the small variation in $E$. Here $V=A \Delta t$ is the integration volume, $N$ is the number of particles in $V$ and $\av{\rho}=mN/V$ is the average energy density of dark matter. We have assumed that the situation is statistically homogeneous and isotropic and evolves little on the timescales it takes for light to cross the interval, so that we can replace the integral along the light ray with a spatial integral \cite{light_stat1, light_stat2, Lavinto:2013exa}. When converting back into an integral, we have smeared the contribution from the small fraction of volume inside the particles into an average contribution over the light path, again assuming that changes are slow. Note how the small average density emerges from the large spiky local density for the first term, which is the same as in geometrical optics. However, there is no such suppression for the second factor of density, which gives a correction of the order $M^2/E^2$.

\para{Correction term 2.}

The second post-geometrical term in \re{thetaint} involves the integral
\bea \label{nabla2rho}
  \frac{1}{4 \Mpl^2} \int_{\Delta\lambda} \rmd\lambda \Box(\rho-p) &=& \frac{1}{4 \Mpl^2} \int\rmd\lambda \d^{ij} \pat_i \pat_j (\rho - p ) \el
  &=& \frac{A^{-1} m}{4 \Mpl^2} \sum_n \int_V \rmd^3 r E^{-1} \d^{ij} \pat_i \pat_j W(r_n) \el
  &=& \frac{A^{-1} m}{4 \Mpl^2} \sum_n \int_V \rmd^3 r E^{-1} \d^{ij} \pat_i \pat_j W(r) \el
  &=& \frac{A^{-1} N m}{4 \Mpl^2} \int_V \rmd^3 r E^{-2} \pat_r W(r) \pat_r E \ ,
\eea
where we have taken into account $p\ll\rho$. As spacetime inside each particle is static, $E\propto|g_{00}|^{-1/2}$. With $u^\a=|g_{00}|^{-1/2}$, we have $\dot u_r=\ha\pat_r\ln|g_{00}|\approx\ha\pat_r|g_{00}|$, which according to \re{udot} is equal to $-\rho^{-1}\pat_r p$, so $\pat_r E=E \rho^{-1} \pat_r p$. The pressure gradient is given by the Tolman--Oppenheimer--Volkov equation (dropping the subscript $n$),
\bea \label{pder}
  \frac{\partial p}{\partial r} &=& - \GN \frac{ (\rho+p) ( \tilde m + 4\pi r^3 p) }{r (r-2\GN \tilde m)} = - \frac{1}{8\pi\Mpl^2} \frac{ \rho \tilde m }{r^2} \ ,
\eea
where $\tilde m(r)\equiv\int_0^r \rmd r' 4\pi r'^2 \rho$, and in the second equality we have taken into account that for a Gaussian density distribution and $m\ll\Mpl$, the Newtonian limit is a good approximation. Therefore
\bea \label{Eder}
  \frac{\partial E}{\partial r} &=& - \frac{E}{8\pi\Mpl^2} \frac{\tilde m}{r^2} \ ,
\eea
Applying \re{Eder}, the integral \re{nabla2rho} becomes
\bea \label{nabla2rho2}
  \int_{\Delta\lambda} \rmd\lambda \frac{1}{4 \Mpl^2} \Box(\rho-p) &=& - \frac{A^{-1} N m E^{-1}}{8\Mpl^4} \int_0^\infty \rmd r \tilde m(r) \pat_r W(r) \el
  &=& \frac{A^{-1} N m E^{-1}}{8\Mpl^4} \int_0^\infty \rmd r \pat_r \tilde m(r) W(r) \el
  &=& \int_{\Delta\lambda} \rmd\lambda \frac{\rho^2}{8\Mpl^2} \ ,
\eea
where we have neglected the subleading variation of $E$. This term cancels half of the correction term in \re{rho2}.

\para{Correction term 3.}

The third post-geometrical term in \re{thetaint} involves an integral over \re{AA}. It contains $\dot k^\a$, which we can write as
\bea \label{kdot}
  \dot k_\a &=& u^\b \nabla_\b k_\a \el
  &=& - \pat_\a E - E v \left( \frac{1}{3} \theta e_\a + \sigma_{\a\b} e^\b + \omega_{\a\b} e^\b - \dot u \cdot e \, u_\a \right) \el
  &=& - \pat_\a E - E v \rho^{-1} e^\b \pat_\b p \, u_\a \ ,
\eea
where we have used $\nabla_\b k_\a=\nabla_\a k_\b$, $E=-u \cdot k$, the decomposition \re{udec}, and on the third line we have used \re{udot}, and taken into account $p\ll\rho$ and that inside the particles $\theta=0$ and $\sigma_{\a\b}=\omega_{\a\b}=0$. Using \re{kdot} in \re{AA}, the integral over $P_{\a\b}P^{\a\b}$ reads
\bea \label{AAint}
   && \int\rmd\lambda \frac{1}{\Mpl^4 E^2 v^2} \left[ - \frac{1}{16} ( e^\a\pat_\a \rho)^2 + \frac{1}{8} \left( 1+ \frac{4 \Mpl^2 E^2}{\rho} \right) e^\a\pat_\a \rho \, e^\b \pat_\b p \right. \el
  && + \frac{1}{16} \left( 3 - 24 \frac{\Mpl^2 E^2}{\rho} \right) ( e^\a \pat_\a p )^2 + \frac{1}{8} \pat_\a (\rho-p) \pat^\a (\rho-p) \el
  && \left. - \Mpl^2 E \pat_\a E \pat^\a (\rho-p) + \Mpl^2 \rho \pat_\a E \pat^\a E \right] \ .
\eea
The directional derivatives in the first three terms again become sums over dot products. For each particle $n$, we have $e^\a \pat_\a = \d_{ij} e^i (r^j-r_n^j)/r_n \pat_{r_n}$. In the sum over particles, the only pairs that give non-zero results are those where the angle is the same, so $\sum_{m,n} \d_{ij} e^i (r^j-r_m^j)/r_m \d_{kl} e^k (r^l-r_n^l)/r_n=\frac{1}{3}\sum_n$, again assuming the change in light ray direction is small or symmetric around the centre of each particle. We get 
\bea \label{AAint2}
   && \int\rmd\lambda \sum_n \frac{1}{\Mpl^4 E^2 v^2} \left[ \frac{5}{48} ( \pat_{r_n} \rho)^2 + \frac{1}{24} \left( - 5 + \frac{4 \Mpl^2 E^2}{\rho} \right) \pat_{r_n} \rho \, \pat_{r_n} p \right. \el
  && \left. + \frac{1}{16} \left( 3 - 8 \frac{\Mpl^2 E^2}{\rho} \right) ( \pat_{r_n} p )^2 - \Mpl^2 E \pat_{r_n} E \pat_{r_n} (\rho-p) + \Mpl^2 \rho ( \pat_{r_n} E)^2 \right] \ .
\eea
Using $\pat_{r_n} E=E \rho^{-1} \pat_{r_n} p$, \re{AAint2} becomes
\bea \label{AAint3}
   && \int\rmd\lambda \sum_n \frac{1}{\Mpl^4 E^2 v^2} \left[ \frac{5}{48} ( \pat_{r_n} \rho)^2 - \frac{1}{24} \left( 5 + \frac{20 \Mpl^2 E^2}{\rho} \right) \pat_{r_n} \rho \, \pat_{r_n} p \right. \el
  && \left. + \frac{1}{16} \left( 3 + \frac{24 \Mpl^2 E^2}{\rho} \right) ( \pat_{r_n} p )^2 \right] \ ,
\eea
where the pressure gradient is given in \re{pder}. We approximate the common prefactor $v^2$ as unity. If $v$ is appreciably different from unity, it will vary inside the particle, but this is relevant only when the light wave nearly has too little energy to cross the particle. Taking into account the variation would not change the order of magnitude of the corrections, but the contribution from the variation and the leading term would contribute to the subleading terms. Somewhat inconsistently, we do take into account the variation of $E$ in the $(\pat_{r_n} \rho)^2$ term, as it can be readily done and affects the subleading terms. We expand $E^{-2}\approx E_0^{-2}(1-2rE_0^{-1}\pat_r E)$, where $E_0$ is the energy at $r_n=0$, giving the extra term $-\frac{5}{24} \rho^{-1} ( \pat_{r_n} \rho)^2r\pat_r p$ in the integral \re{AAint3} compared to the case when the variation in $E$ is neglected. (The contribution of this term turns out to be $-2$ times the contribution of the term $-\frac{5}{24} \pat_{r_n} \rho \pat_{r_n} p$.) The terms in \re{AAint2} reduce to volume integrals involving factors of $\rho$, $\pat_r\rho$, $\tilde m$ and $r$. Using the Gaussian density distribution \re{rho} and the pressure gradient \re{pder}, we have
\bea \label{numint}
  \int\rmd^3 r (\pat_r \rho)^2 &=& \frac{m}{16\pi^{5/2}} m^2 \rho_0 \el
  \int\rmd^3 r \pat_r \rho \pat_r p &=& \frac{m}{54\sqrt{3}\pi} \frac{\rho_0^2}{\Mpl^2} \el
  \int\rmd^3 r \rho^{-1} \pat_r \rho \pat_r p &=& \frac{m}{12\sqrt{\pi}} \frac{\rho_0}{\Mpl^2} \el
  \int\rmd^3 r \rho^{-1} (\pat_r \rho)^2 r \pat_r p &=& - \frac{m}{27\sqrt{3}\pi} \frac{\rho_0^2}{\Mpl^2} \el
  \int\rmd^3 r (\pat_r p)^2 &\approx& 3\times10^{-9} m \frac{m^4}{\Mpl^4} m^2 \rho_0 \el
  \int\rmd^3 r \rho^{-1} (\pat_r p)^2 &\approx& 4\times10^{-5} m \frac{m^2}{\Mpl^2} \frac{\rho_0}{\Mpl^2} \ .
\eea
The precise numerical values depend on the form of the density distribution, but the orders of magnitude are simple to understand. Over the integral, the radial derivative gives a factor of $\lc^{-1}$ as that is the scale of variation of the density, so $\int\rmd^3 r (\pat_r \rho)^2\sim\lc^{-2} m \rho_0\sim10^{-2} m^3 \rho_0$. The pressure gradient is suppressed by the factor $m^2/\Mpl^2$ compared to the density gradient, so terms proportional to $(\pat_r p)^2$ are negligible. For a steeper density profile, the integrals involving $\pat_r\rho$ would be larger; for a step function the integral over $(\pat_r\rho)^2$ diverges. In the integral over the light rays, one factor of $m$ combines with $N/V$ to become the average density $\av{\rho}$ as before. Overall, \re{AAint3} becomes
\bea \label{AAfinal}
\int\rmd\lambda E^2 \frac{\av{\rho}}{\Mpl^2} \left[ \frac{5}{768\pi^{5/2}} \frac{m^2}{E^2} \frac{\rho_0}{\Mpl^2 E^2} - \frac{5}{72\sqrt{\pi}} \frac{\rho_0}{\Mpl^2 E^2} + \frac{5}{1296\sqrt{3}\pi} \left( \frac{\rho_0}{\Mpl^2 E^2} \right)^2 \right] \ .
\eea

\para{The total correction.}

In summary, gathering the terms \re{rho2}, \re{nabla2rho2} and \re{AAfinal}, the integral \re{thetaint} is
\bea \label{thetaint2}
  \int\rmd\lambda \thetat &=& \int\rmd\lambda \int\rmd\lambda' \left\{ - \frac{1}{2}\tilde{\theta}^2 - 2\sigmat^2 - E^2 \frac{\av{\rho}}{\Mpl^2} \left[ 1 + \frac{5}{8192\pi^{13/2}} \frac{m^6}{\Mpl^2 E^4} \right. \right. \el
  && \left. \left. - \frac{13}{144\sqrt{\pi}} \frac{\rho_0}{\Mpl^2 E^2} + \frac{5}{1296\sqrt{3}\pi} \left( \frac{\rho_0}{\Mpl^2 E^2} \right)^2 \right] \right\} \ .
\eea
The maximum density is $\sqrt{2/(9\pi)}\rho_0$, so the maximum induced mass squared is $M^2=\rho_0/(\sqrt{18\pi}\Mpl^2)$. If $E^2$ is smaller than this, the wave cannot propagate through the particle. For this limiting energy, the last two terms in \re{thetaint2} reach their maximum absolute value 0.4 and 0.04, respectively. Although the numerical prefactors are sensitive to the density profile, the general picture is that these correction terms are small until the energy falls close to the curvature scale, when our approximation is not necessarily valid any more. Once the curvature-induced mass exceeds the energy, light cannot propagate inside the particles, and our plane wave approximation is certainly not valid.

These last two correction terms determined by the curvature scale are in line with the naive estimates discussed in the introduction. In contrast, the first correction term in \re{thetaint2} is given not by the curvature, but by the square of the derivative of the curvature, and is hence enhanced by the factor $m^2/E^2$. This term can give a large correction to the angular diameter distance even when the curvature is much smaller than the photon energy, and the dispersion relation of light is close to the vacuum case. The relative correction is unity at $E=0.02 (m^3/\Mpl)^{1/2}$. (Note that in this limit, the derivative of the curvature remains small relative to $E^3$.) Expressed another way, for photons of energy $E$, this correction is significant if $m\gtrsim10 (\Mpl E^2)^{1/3}$. For the longest radio waves observable through the Earth's ionosphere, this gives $m\gtrsim100$ keV, and for CMB photons we get $m\gtrsim100$ MeV. If correct, this is a strong result: we can rule out dark matter masses $10^4$ GeV $\gtrsim m\gtrsim100$ MeV from the absence of a signal in the CMB. Although the precise limit on the mass changes with the density distribution inside the particle, the energy dependence does not. If we assumed that light can pass through electrons and nuclei in the same way as through dark matter, they would have negligible effect on the CMB, but not necessarily on radio waves.

\para{From $\thetat$ to $D_A$.}

In terms of the angular diameter distance $D_A$, \re{thetaint2} corresponds to (see \eg \cite{light_stat1, light_stat2}), dropping those correction terms that are important only near the curvature limit, and neglecting the null shear (which is small in the real universe),
\bea \label{DA}
   D_A'' &=& - 4\pi\GN \av{\rho} \left[ 1 + \a (E/E_\text{o})^{-4} \right] D_A \ ,
\eea
where $\a\equiv\frac{5}{8192\pi^{13/2}} \frac{m^6}{\Mpl^2 E_\text{o}^4}$. In contrast to geometrical optics, the angular diameter distance depends on energy. Switching from the affine parameter $\lambda$ to the redshift $z$ using $\frac{\rmd}{\rmd\lambda}=-(1+z)H(z)E\frac{\rmd}{\rmd z}$, where $H(z)$ is the Hubble parameter, we obtain
\bea \label{DAz}
  H(z) \frac{\rmd}{\rmd z} \left[ (1+z)^2 H(z) \frac{\rmd D_A}{\rmd z} \right] &=& - 4\pi\GN \av{\rho} \left[ 1 + \a (1+z)^{-4} \right] D_A \ .
\eea
Because $\av{\rho}\propto (1+z)^3$, the new contribution to the source term in the angular diameter distance grows like $(1+z)^{-1}$, and becomes important at late times. The redshift dependence is the same as for dark energy with equation of state $p_\text{de}=-\frac{4}{3}\rho_\text{de}$. However, the sign $\a>0$ in \re{DAz} corresponds to $\rho_\text{de}<0$. Also, it is impossible to get a correction term like this from a matter source in a FLRW universe, because matter would also modify the mapping between $\lambda$ and $z$ given by $H(z)$, not just the source term, unlike here. This is the reverse of corrections from backreaction, which change the $\lambda-z$ mapping, but not the source term \cite{light_stat1, light_stat2, Bull:2012zx, Lavinto:2013exa}. A value $\a\gtrsim1$ on a given wavelength would significantly affect $D_A$, and measurements of distances using light with different wavelength would give discrepant results when interpreted without the correction term. As $\a>0$, the correction leads to a shorter distance, so this effect cannot mimic a cosmological constant. Nor can it explain the tension between the determination of $H_0$ (\ie the inverse of the normalisation of the distance) from the CMB and local measurements of $H_0$ (which are only weakly dependent on a correction like this) \cite{Verde:2019ivm}, because the inferred value of $H_0^{-1}$ from the CMB is larger. (If $\a$ had the opposite sign, it could have such an effect, and would for the right value even change $D_A$ in a manner similar to the cosmological constant.)

We have considered dark matter only, but $\av{\rho}$ usually also includes baryonic matter. (As the full term is $\av{\rho}+\av{p}$, the cosmological constant does not contribute.) If we assume that light cannot propagate through nuclei due to electromagnetic interactions, they do not contribute to the emergence of the average density from the spiky local density. This leads to reduced density along the line of sight, and hence increased distance. In the $\Lambda$CDM model the Planck mean parameters are $\Omega_\text{m}=\Omega_\text{dm}+\Omega_\text{b}=0.315$, $\Omega_\Lambda=1-\Omega_\text{m}$, and $\Omega_\text{b}=0.0224/h^2$, where $h=H_0/(100 \text{km/s/Mpc)}$ \cite{Aghanim:2018eyx}. We can roughly estimate the effect of not propagating through baryons by dropping the baryonic contribution to the angular diameter distance integral $D_A= H_0^{-1} (1+z)^{-1} \int_0^z \frac{\rmd z'}{\sqrt{ \Omega_\Lambda + (\Omega_\text{dm}+\Omega_\text{b}) (1+z')^3 }}$. As the denominator in the integral is then smaller than in the usual $\Lambda$CDM case, the distance is longer. If we fit the $\Lambda$CDM model (with baryons included) to the CMB, the inferred value of $H_0$ will be smaller to compensate. If we adopt the local measurement $H_0=73.2\pm1.3$ km/s/Mpc \cite{Riess:2020fzl} as the real value, the inferred value is $69.0\pm1.2$ km/s/Mpc, closer to the value $67.4\pm0.5$ km/s/Mpc reported by Planck. Other determinations of $H_0$ involving the integral over the light ray would also be affected differently. In any case, this estimate is rather naive, and the issue of light propagation inside charged particles should be considered in more detail.

Possible violation of the distance duality between the angular diameter distance and the luminosity distance could also be important, as the duality is a key input in drawing conclusions about the compatibility of CMB and baryon acoustic oscillation observations and observations of type Ia supernovae. In appendix \ref{sec:duality} we show that although the deviation equation of two light rays receives a large post-geometrical correction, the reciprocity relation and hence distance duality remains unaffected.

\subsection{Discussion} \label{sec:disc}

\para{Assumptions about light and matter.}

Let us consider the approximations we have made concerning light propagation and modelling of matter. We have assumed that the scales over which amplitude and polarisation of the light wave change are large compared to the wavelength, while spacetime curvature can be large, and that the electromagnetic field has the form of a local plane wave. This is only consistent if the energy flux and anisotropic stress can be neglected and the polarisation is orthogonal to the observer four-velocity. The first two conditions are reasonable, the last is more problematic, although differences in source and observer velocity are small in practice.

In the case of the amplitude, if the induced mass is close to the photon energy, $M\sim E$, the dispersion relation is strongly modified, so we could expect that the light wavefront does not remain straight, and the wave becomes choppy on small scales as it breaks on the dark matter particles. For general values of $M/E$, we can estimate the consistency of assuming that the amplitude changes slowly from \re{pn}, which gives $a'= - \ha a \nabla_\a k^\a$. (Note that this equation remains valid even outside our approximation as long as photon number is conserved.) So $a\propto 1/D_A$ changes slowly. As for the polarisation, contracting \re{pol2} with $k^\a$ and applying the decomposition \re{udec} gives the estimate $f_\a'\sim (E m^3)/\Mpl^2\sim (E M^2)/m$. The second derivative of the polarisation in the fourth term in \re{eomexp} is $\sim E^{-2} f_\a''\sim E^{-1} m f_\a'\sim M^2$. So in the observationally allowed regime where the leading correction to $D_A$ is below unity it is formally much smaller than the wavenumber $|\vec k|\approx E$, which sets the expansion scale. Of course both $k^2$ and the curvature term are smaller than $E^2$ as well (in the geometrical optics approximation, the leading term $k^2$ is zero), so this issue should be checked in detail.

In any case, these order of magnitude estimates can at best support consistency, not establish correctness. A related issue is that our approximation may not treat regimes of intermediate curvature between the curvature spikes and empty space correctly. If the polarisation and/or the amplitude also vary rapidly, this could naively be expected to push the situation even further away from the range of validity geometrical optics approximation, though correction terms could cancel. The validity of these approximations related to light propagation can in principle be straightforwardly checked with numerical solutions of the Maxwell equation in curved spacetime, given a matter distribution. This brings us to the assumptions related to the modelling of matter.

We have assumed that dark matter particle mass can be treated as being spread in a classical Gaussian distribution, which is related to spacetime curvature via the Einstein equation. While the particles are inherently quantum mechanical, the spatial distances and curvatures we consider are far below the Planck scale, so this should be a reasonable approximation. If that were not the case, then the quantum mechanical nature of the particles and/or their gravity would play a role, which would push the situation even further from the range of validity of the usual treatment. A key assumption is that the width of the Gaussian is given by the Compton wavelength, and the numerical limits we obtain depend on this choice (and on the overall shape of the mass distribution). If the width of the Gaussian is $l$ instead of $\lc$, the density changes by the factor $(\lc/l)^3$, and the leading correction to the angular diameter distance picks an extra factor of $(\lc/l)^2$ from two derivatives. For example, if we used the de Broglie wavelength $2\pi/(m v_\text{dm})$ instead, where $v_\text{dm}$ dark matter velocity, the mass related to the curvature limit would increase by the factor $v_\text{dm}^{-3/4}\sim10^2$ (taking $v_\text{dm}\sim10^{-3}$), to $m\sim10^6$ GeV. The limiting mass for the leading correction to the angular diameter distance would increase by the factor $v_\text{dm}^{-5/6}\sim10^2$, to $\sim 10$ GeV.

The mass distribution is determined by the wave function (or more generally the density matrix) of dark matter particles. The width of a free particle wavepacket grows linearly in time, and would quickly grow to make the curvature spikes negligible. However, cosmological dark matter particles constantly interact via gravity --and many candidates also via other interactions-- with light, neutrinos, baryonic matter and each other. Dark matter particles decohere via these interactions \cite{Joos:1984uk, Tegmark:1993yn, Allali:2020ttz, Allali:2020shm, Allali:2021puy}, and should also localise, as otherwise their wavefunctions would have spread over billions of light years by today. However, if the particles remain delocalised over scales much larger than the inverse mass, the limits on the mass we have found become too weak to make any difference.

\section{Conclusions} \label{sec:conclusion}

\para{Limits on dark matter from light propagation.}

In general relativity, light travel on null geodesics is an approximation valid in the geometrical optics limit when wavelength is small compared to the spacetime curvature radius and the scale over which the amplitude and the polarisation change. We have introduced a post-geometrical approximation that includes spacetime curvature, and derived the expressions for the redshift and angular diameter distance. We find that inside matter photons acquire a gravity-induced mass $M$ of the order of the curvature scale. Light paths are not null, and the gradient of the matter density (and pressure) pushes photons off the geodesic path. If $M>E$, light cannot propagate inside the particles. We approximate dark matter particles of mass $m$ as Gaussian wavepackets with width equal to the Compton wavelength $\lc$. We find that the correction to the redshift is negligible. 

In the angular diameter distance $D_A$, the integral of the spiky density over the light ray reproduces the usual effect of the average density. However, we also find a post-geometrical correction that is enhanced by the factor $m^2/E^2$ over the naive expectation that the effects are suppressed by $M^2/E^2$. The reason is that not only the density but also its derivatives enter into $D_A$, and the density changes on the scale $\lc\sim1/m$ inside the particle. The origin of the derivatives is the geodesic deviation vector $k^\b\nabla_\b k^\a$, which vanishes in geometrical optics, but here is proportional to the gradient of the density. The correction to the angular diameter distance is of order unity or larger for $m\gtrsim10(\Mpl E^2)^{1/3}$. For CMB photons $E\sim(1+z)10^{-3}$ eV, and the angular diameter distance to the CMB is strongly affected for $m\gtrsim 100$ MeV. The curvature-suppressed corrections would give the much weaker limit $m\gtrsim10\sqrt{\Mpl E}\sim10^4$ GeV.

As the correction term to the angular diameter distance depends on photon energy, observations made at different wavelengths could yield discrepant results if this term is not taken into account. The effect only makes $D_A$ smaller, so it cannot mimic the effect of a cosmological constant nor explain the $H_0$ tension between the CMB and other probes, and can only make the fit to observations worse.

We find that polarisation undergoes gravity-driven rotation which could provide an interesting observational signature. It would be interesting to look at the effect not only on the angular diameter distance (\ie beam convergence), but also on lensing (\ie beam deformation), studied for extended beams in the geometrical optics approximation in \cite{Fleury:2017owg, Fleury:2018cro, Fleury:2018odh}. In addition to considering in more detail the case $M\sim E$, when our approximation may not be valid, it would be interesting to go beyond to the case $M>E$, when light cannot propagate inside dark matter, which falls outside the local plane wave approximation.

If the results hold, dark matter masses $10^4$ GeV $\gtrsim m\gtrsim100$ MeV are ruled out by CMB observations, and observations of $D_A$ using longer wavelengths, such as the 21 cm line, offer the prospect of tightening this bound by a few orders of magnitude. This may offer a novel way to probe the microscopic nature of dark matter via observations of light. However, a more detailed study of the validity of the assumptions we have made about light propagation and modelling of dark matter is required before drawing such conclusions.

\acknowledgments

We thank Pierre Fleury for helpful correspondence and the anonymous referee for useful criticism. Sofie Marie Koksbang is funded by the Carlsberg Foundation.

\appendix

\section{Calculating $D_A$} \label{sec:area}

In this appendix we relate the beam area expansion rate to the angular diameter distance $D_A$, generalising the geometrical optics calculation (see \eg \cite{Lensing} pages 104-115). Let $\d x^\a$ be a separation vector connecting two rays in an infinitesimal light beam. Note that in order for the difference in the position of two rays to be a vector, the rays have to belong to the same tangent space, \ie the curvature has to be small. When $E\sim M$, the curvature is large, and the validity of the treatment is questionable, whereas for $E\gg M$ the curvature is small. As seen in \sec{sec:corr}, in order to reproduce the average cosmological density of matter in agreement with observations we have to consider beams that are larger than the separation of dark matter particles. We simply assume that we can sum up the effects of nearby light rays, but it would be interesting to consider rigorously how the infinitesimal beams can be added, and how the tangent spaces orthogonal to the light rays mesh together.

By definition, $\d x^\a$ is dragged along the light rays, so its Lie derivative with respect to the tangent vector $k^\a$ vanishes,
\begin{align} \label{Liex}
  k^{\b}\nabla_{\b}\delta x^{\a} - \delta x^{\b}\nabla_{\b}k^{\a}  = 0 \ .
\end{align}
Let $\tilde U^\a$ be a timelike vector dragged along the light rays, so that its Lie derivative with respect to $k^\a$ also vanishes,
\begin{align} \label{Lieu}
  k^{\b}\nabla_{\b} \tilde U^{\a} - \tilde U^{\b}\nabla_{\b} k^{\a} = 0 \ .
\end{align}
The unit normalised four-velocity parallel to $\tilde U^\a$ is $U^\a=N \tilde U^\a$ (here $N\equiv1/\sqrt{-\tilde U \cdot \tilde U}$, so that $U\cdot U=-1$):
\bea \label{kdecU}
  k^\a &=& E_{(U)} ( U^\a + v_{(U)} e_{(U)}^\a ) \ ,
\eea
where $E_{(U)}=-U \cdot k$, $U \cdot e_{(U)}=0$, $e_{(U)} \cdot e_{(U)} = 1$, and $v_{(U)}\equiv\sqrt{1+k^2/E_{(U)}^2}$.
Along the light ray, $\d x^\a$ will not stay orthogonal to $k^\a$ even if it was initially so. We therefore introduce the projection tensor orthogonal both to $k^\a$ and $U^\a$, analogous to \re{htt},
\bea \label{httU}
  \htt_{\a\b}^{(U)} &\equiv& g_{\a\b} + U_\a U_\b - e_{(U)\a} e_{(U)\b} \el
  &=& g_{\a\b} + \frac{k^2}{E_{(U)}^2+k^2} U_\a U_\b + \frac{2 E_{(U)}}{E_{(U)}^2+k^2} U_{(\a} k_{\b)} - \frac{1}{E_{(U)}^2+k^2} k_\a k_\b \ .
\eea
The orthogonally projected separation vector is $h_\a\equiv \htt_{\a\b} \d x^\b\equiv h \hat h_\a$, where $\hat h^{\a} \hat h_\a = 1$. The norm $h>0$ gives the distance between light rays in the bundle. Let us consider its evolution along the light ray,
\bea \label{thetatder}
  \frac{\rmd}{\rmd\lambda} h^2 &=& 2 h^\a k^\c \nabla_\c ( \htt_{\a\b}^{(U)} \d x^\b ) \el
  &=& 2 h_\b k^\c \nabla_\c \d x^\b + 2 h^\a \d x^\b k^\c \nabla_\c \htt_{\a\b}^{(U)} \el
  &=& 2 h_\b \d x^\c \nabla_\c k^\b + 2 h^\a \d x^\b k^\c \nabla_\c \htt_{\a\b}^{(U)} \el
  &=& 2 h^\a h^\b \nabla_\b k_\a + 2 \frac{k^2 \d x \cdot U + E_{(U)} \d x \cdot k}{E_{(U)}^2+k^2} h^\a ( k^{\b}\nabla_{\b} U_\a - U^{\b}\nabla_{\b} k_\a ) \el
  &=& 2 h^\a h^\b \nabla_\b k_\a + 2 \frac{k^2 \d x \cdot U + E_{(U)} \d x \cdot k}{E_{(U)}^2+k^2} h^\a \tilde U_\a k^{\b}\nabla_{\b} N \el
  &=& 2 h^\a h^\b \nabla_\b k_\a \ ,
\eea
where we have on the third line used \re{Liex}, on the fourth line written \mbox{$\d x^\c=h^\c + ( \d^\c{}_\a - \htt^\c{}_\a ) \d x^\a$} and used the definition \re{httU}, on the next to last line written $U^\a=N \tilde U^\a$ and used the vanishing of the Lie derivative \re{Lieu}, and on the last line taken into account that $h^\a$ and $U^\a$ are orthogonal.

Analogously to \re{nablakdec}, we decompose $\nabla_\b k_\a$ into parts projected in the direction of and orthogonal to $\htt_{\a\b}^{(U)}$ as
\bea \label{nablakdecU}
  \nabla_\b k_\a = \frac{1}{2} \thetat^{(U)} \htt_{\a\b}^{(U)} + \sigmat_{\a\b}^{(U)} + P_{\a\b}^{(U)} \ ,
\eea
where the area expansion rate is $\thetat^{(U)}\equiv \htt^{(U) \a\b} \nabla_\b k_\a$, the light shear is $\sigmat_{\a\b}^{(U)}=\htt^{(U)}_{\a\c} \htt_{\b\d} \nabla^\d k^\c-\ha \htt_{\a\b}^{(U)} \thetat^{(U)}$, and $P_{\a\b}^{(U)}\equiv ( g_{\a\c} g_{\b\d} - \htt_{\a\c}^{(U)} \htt_{\b\d}^{(U)} ) \nabla^\d k^\c$. In geometrical optics, $P_{\a\b}^{(U)}$ is orthogonal to $k^\a$ and can be written as $P_{\a\b}^{(U)}=2 k_{(\a} P_{\b)}^{(U)}$, with $P^{(U)}\cdot k=0$. That is not the case here.

Inserting \re{nablakdecU} into \re{thetatder}, we obtain
\bea
   \frac{1}{h} \frac{\rmd h}{\rmd\lambda} &=& \frac{1}{2} \thetat^{(U)} + \sigmat_{\a\b}^{(U)} \hat h^\a \hat h^\b \ ,
\eea
from which it follows that $\thetat=\frac{1}{A}\frac{\rmd A}{\rmd\lambda}$ describes the relative rate of change of the beam cross section area $A$. Since the angular diameter distance is defined as $D_A^2 = A/\Omega_\text{o}$, where $\Omega_\text{o}$ is the solid angle at observation, we obtain the usual relation
\bea \label{DAthetat}
  D_A\propto \exp \left(\frac{1}{2} \int\rmd\lambda \thetat^{(U)} \right) \ .
\eea  

Taking the trace of \re{nablakdecU}, we can split the area expansion rate as
\bea \label{thetatU}
  \thetat^{(U)} &=& \nabla_\a k^{\a} - P^{(U)\a}{}_\a \ .
\eea
Velocity-dependence of the area expansion rate is contained in the second term (which is zero in geometrical optics). Using \re{httU}, we can write it as
\bea \label{Dtheta}
  - P^{(U)\a}{}_\a &=& \frac{1}{E_{(U)}^2+k^2} \left( k^2 U^\a U^\b \nabla_\b k_\a + E_{(U)} U^\a \nabla_\a k^2 - \ha k^\a \nabla_\a k^2 \right) \ .
\eea
We can rewrite the first term using the identity (which it is straightforward to verify)
\bea \label{DEU}
   E_{(U)} U^\a U^\b \nabla_\b k_\a &=& k^\a \nabla_\a E_{(U)} + U^\a \nabla_\a k^2 \ .
\eea
Inserting \re{DEU} into \re{Dtheta}, we obtain
\bea \label{Dtheta2}
  - P^{(U)\a}{}_\a &=& \frac{k^2}{E_{(U)} ( E_{(U)}^2+k^2 )} k^\a \nabla_\a E_{(U)} +  \frac{1}{E_{(U)}} U^\a \nabla_\a k^2 - \ha  \frac{1}{E_{(U)}^2+k^2} k^\a \nabla_\a k^2 \el
  &=& \ha \frac{\rmd}{\rmd\lambda} \left( \ln\frac{E_{(U)}^2}{E_{(U)}^2+k^2} \right) + \frac{1}{E_{(U)}} U^\a \nabla_\a k^2 \ .
\eea
When integrating from the source to the observer, the first term gives zero if we assume that they are not coincident with a dark matter particle, so that $k^2=0$. We can decompose $U^\a$ as
\bea
  U^\a = \c ( u^\a + s^\a ) \ ,
\eea
where $\c\equiv -u\cdot U$, $u\cdot s=0$. We thus have
\bea
  - \int\rmd\lambda P^{(U)\a}{}_\a &=& - \int\rmd\lambda \frac{1}{2 \Mpl^2 E_{(U)}} \c s^\a \pat_\a ( \rho - p ) \ ,
\eea
where we have used \re{null2} and the fact that the density and pressure are independent of the time defined by $u^\a$ (but not of the time defined by $U^\a$). As this term is linear in the density, it is parametrically suppressed compared to the other terms we have found that are quadratic in density. However, it is also suppressed by a sum over directions. As in \sec{sec:DA}, the directional derivative becomes a sum over dot products, in this case $\d_{ij} (r^i-r^i_n) s^j/r_n$. As the turning of $U^\a$ (and hence $s^i$) along the light ray has no preferred spatial direction with respect to $u^\a$ apart from $e^\a$, and the term is linear in the cosine of the angle, it vanishes.

Therefore the velocity-dependent term $P^{(U)\a}{}_\a$ can be neglected when calculating $D_A$, and it is sufficient to consider the term $\nabla_\a k^\a$, which is independent of velocity.

\section{Distance duality} \label{sec:duality}

In this appendix, we show that the reciprocity relation and hence distance duality $D_L=(1+z)^2 D_A$, where $D_L$ is the luminosity distance, are unaffected by the correction terms in our post-geometrical optics approximation. We follow the proof for the reciprocity relation in the geometrical optics approximation outlined in \eg \cite{Lensing} (pages 110-115), \cite{RelCosmo} (pages 165-166).

Consider two beams between source and observer, one that converges at the observer and another that converges at the source. We denote a deviation vector in the beam that converges at the source by 1 and a deviation vector in the beam that converges in the observer by 2. The two beams have common light rays. Consider the following quantity along such a ray:
\bea
  \delta x^{\a}_1 k^{\b}\nabla_\b\delta x_{2\a} - \d x^{\a}_2 k^{\b}\nabla_\b\delta x_{1\a} &=& ( \delta x^{\a}_2 \delta x^\b_1 - \d x^{\a}_1 \delta x^\b_2 ) \nabla_\b k_\a = 0 \ ,
\eea
where in the first equality we have used the vanishing of the Lie derivative \re{Liex}, and the second equality follows from the fact that $\nabla_\b k_\a$ is symmetric. Given the equality $\delta x^{\a}_1 k^{\b}\nabla_\b\delta x_{2\a} = \d x^{\a}_2 k^{\b}\nabla_\b\delta x_{1\a}$, the proof of the reciprocity relation between the area and subtended angle at observer and source, and hence distance duality, follows in the same way as in geometrical optics. Unlike in geometrical optics, the screen area is not independent of observer velocity where $k^\a$ is not null. However, this would only make a difference if a significant density of matter overlapped the observer, which is not the case for particle dark matter (at least of the non-fuzzy variety where the density at particle location is large enough to make a difference).

\bibliographystyle{JHEP}
\bibliography{disc}

\providecommand{\href}[2]{#2}\begingroup\raggedright\begin{thebibliography}{10}

\bibitem{PhoneBook}
C.~Misner, K.~Thorne and J.~Wheeler, \emph{Gravitation}.
\newblock Princeton University Press, 1973.

\bibitem{Lensing}
P.~Schneider, J.~Ehlers and E.~Falco, \emph{Gravitational Lenses}.
\newblock Astronomy and Astrophysics Library. Springer New York, 1992.

\bibitem{WaveOptics_4}
T.~T. Nakamura and S.~Deguchi, \emph{{Wave Optics in Gravitational Lensing}},
  \href{http://dx.doi.org/10.1143/PTPS.133.137}{\emph{Progress of Theoretical
  Physics Supplement} {\bfseries 133} (01, 1999) 137--153}.

\bibitem{WaveOptics_5}
Y.~Nambu, \emph{{Wave Optics and Image Formation in Gravitational Lensing}},
  \href{http://dx.doi.org/10.1088/1742-6596/410/1/012036}{\emph{J. Phys. Conf.
  Ser.} {\bfseries 410} (2013) 012036},
  [\href{https://arxiv.org/abs/1207.6846}{{\ttfamily 1207.6846}}].

\bibitem{WaveOptics_2}
S.~G. Turyshev and V.~T. Toth, \emph{{Diffraction of electromagnetic waves in
  the gravitational field of the Sun}},
  \href{http://dx.doi.org/10.1103/PhysRevD.96.024008}{\emph{Phys. Rev. D}
  {\bfseries 96} (2017) 024008},
  [\href{https://arxiv.org/abs/1704.06824}{{\ttfamily 1704.06824}}].

\bibitem{WaveOptics_3}
S.~G. Turyshev and V.~T. Toth, \emph{{Diffraction of electromagnetic waves by
  an extended gravitational lens}},
  \href{http://dx.doi.org/10.1103/PhysRevD.103.064076}{\emph{Phys. Rev. D}
  {\bfseries 103} (2021) 064076},
  [\href{https://arxiv.org/abs/2102.03891}{{\ttfamily 2102.03891}}].

\bibitem{Turyshev:2021eoh}
S.~G. Turyshev and V.~T. Toth, \emph{{Gravitational lensing by an extended mass
  distribution}},
  \href{http://dx.doi.org/10.1103/PhysRevD.104.044013}{\emph{Phys. Rev. D}
  {\bfseries 104} (2021) 044013},
  [\href{https://arxiv.org/abs/2106.06696}{{\ttfamily 2106.06696}}].

\bibitem{GW_1}
T.~T. Nakamura, \emph{Gravitational lensing of gravitational waves from
  inspiraling binaries by a point mass lens},
  \href{http://dx.doi.org/10.1103/PhysRevLett.80.1138}{\emph{Phys. Rev. Lett.}
  {\bfseries 80} (Feb, 1998) 1138--1141}.

\bibitem{GW_2}
R.~Takahashi and T.~Nakamura, \emph{{Wave effects in gravitational lensing of
  gravitational waves from chirping binaries}},
  \href{http://dx.doi.org/10.1086/377430}{\emph{Astrophys. J.} {\bfseries 595}
  (2003) 1039--1051}, [\href{https://arxiv.org/abs/astro-ph/0305055}{{\ttfamily
  astro-ph/0305055}}].

\bibitem{GW_3}
J.-P. Macquart, \emph{{Scattering of gravitational radiation: Second order
  moments of the wave amplitude}},
  \href{http://dx.doi.org/10.1051/0004-6361:20034512}{\emph{Astron. Astrophys.}
  {\bfseries 422} (2004) 761--775},
  [\href{https://arxiv.org/abs/astro-ph/0402661}{{\ttfamily
  astro-ph/0402661}}].

\bibitem{WaveOptics_GW}
R.~Takahashi, \emph{{Amplitude and phase fluctuations for gravitational waves
  propagating through inhomogeneous mass distribution in the universe}},
  \href{http://dx.doi.org/10.1086/503323}{\emph{Astrophys. J.} {\bfseries 644}
  (2006) 80--85}, [\href{https://arxiv.org/abs/astro-ph/0511517}{{\ttfamily
  astro-ph/0511517}}].

\bibitem{GW_4}
R.~Takahashi, \emph{{Arrival time differences between gravitational waves and
  electromagnetic signals due to gravitational lensing}},
  \href{http://dx.doi.org/10.3847/1538-4357/835/1/103}{\emph{Astrophys. J.}
  {\bfseries 835} (2017) 103},
  [\href{https://arxiv.org/abs/1606.00458}{{\ttfamily 1606.00458}}].

\bibitem{GW_5}
G.~Cusin and M.~Lagos, \emph{{Gravitational wave propagation beyond geometric
  optics}}, \href{http://dx.doi.org/10.1103/PhysRevD.101.044041}{\emph{Phys.
  Rev. D} {\bfseries 101} (2020) 044041},
  [\href{https://arxiv.org/abs/1910.13326}{{\ttfamily 1910.13326}}].

\bibitem{DeWitt:1960fc}
B.~S. DeWitt and R.~W. Brehme, \emph{{Radiation damping in a gravitational
  field}}, \href{http://dx.doi.org/10.1016/0003-4916(60)90030-0}{\emph{Annals
  Phys.} {\bfseries 9} (1960) 220--259}.

\bibitem{MaxwellEFE_1}
F.~Cabral and F.~S.~N. Lobo, \emph{{Electrodynamics and Spacetime Geometry:
  Foundations}},
  \href{http://dx.doi.org/10.1007/s10701-016-0051-6}{\emph{Found. Phys.}
  {\bfseries 47} (2017) 208--228},
  [\href{https://arxiv.org/abs/1602.01492}{{\ttfamily 1602.01492}}].

\bibitem{MaxwellEFE_4}
F.~A. Asenjo and S.~A. Hojman, \emph{{Do electromagnetic waves always propagate
  along null geodesics?}},
  \href{http://dx.doi.org/10.1088/1361-6382/aa8b48}{\emph{Class. Quant. Grav.}
  {\bfseries 34} (2017) 205011},
  [\href{https://arxiv.org/abs/1608.06572}{{\ttfamily 1608.06572}}].

\bibitem{Copi:2020qur}
C.~J. Copi, K.~Pasmatsiou and G.~D. Starkman, \emph{{Scalar and vector tail
  radiation from the interior of the lightcone}},
  \href{http://dx.doi.org/10.1088/1475-7516/2021/01/050}{\emph{JCAP} {\bfseries
  01} (2021) 050}, [\href{https://arxiv.org/abs/2008.13069}{{\ttfamily
  2008.13069}}].

\bibitem{critique_1}
P.~D. Mannheim, \emph{{Critique of the use of geodesics in astrophysics and
  cosmology}},  \href{https://arxiv.org/abs/2105.08556}{{\ttfamily
  2105.08556}}.

\bibitem{HigherOrder_1}
A.~M. Anile, \emph{Geometrical optics in general relativity: A study of the
  higher order corrections},
  \href{http://dx.doi.org/10.1063/1.522946}{\emph{Journal of Mathematical
  Physics} {\bfseries 17} (1976) 576--584}.

\bibitem{Dolan:2018ydp}
S.~R. Dolan, \emph{{Higher-order geometrical optics for electromagnetic waves
  on a curved spacetime}},  \href{https://arxiv.org/abs/1801.02273}{{\ttfamily
  1801.02273}}.

\bibitem{Oancea:2019pgm}
M.~A. Oancea, C.~F. Paganini, J.~Joudioux and L.~Andersson, \emph{{An overview
  of the gravitational spin Hall effect}},
  \href{https://arxiv.org/abs/1904.09963}{{\ttfamily 1904.09963}}.

\bibitem{Oancea:2020khc}
M.~A. Oancea, J.~Joudioux, I.~Y. Dodin, D.~E. Ruiz, C.~F. Paganini and
  L.~Andersson, \emph{{Gravitational spin Hall effect of light}},
  \href{http://dx.doi.org/10.1103/PhysRevD.102.024075}{\emph{Phys. Rev. D}
  {\bfseries 102} (2020) 024075},
  [\href{https://arxiv.org/abs/2003.04553}{{\ttfamily 2003.04553}}].

\bibitem{Frolov:2020uhn}
V.~P. Frolov, \emph{{Maxwell equations in a curved spacetime: Spin optics
  approximation}},
  \href{http://dx.doi.org/10.1103/PhysRevD.102.084013}{\emph{Phys. Rev. D}
  {\bfseries 102} (2020) 084013},
  [\href{https://arxiv.org/abs/2007.03743}{{\ttfamily 2007.03743}}].

\bibitem{Zeldovich:1964}
Y.~B. {Zel'dovich}, \emph{{Observations in a Universe Homogeneous in the
  Mean}}, {\emph{Soviet Astronomy} {\bfseries 8} (Aug., 1964) 13}.

\bibitem{Dashevskii:1965}
V.~M. {Dashevskii} and Y.~B. {Zel'dovich}, \emph{{Propagation of Light in a
  Nonhomogeneous Nonflat Universe II}}, {\emph{Soviet Astronomy} {\bfseries 8}
  (June, 1965) 854}.

\bibitem{Dashevskii:1966}
V.~M. {Dashevskii} and V.~I. {Slysh}, \emph{{On the Propagation of Light in a
  Nonhomogeneous Universe}}, {\emph{Soviet Astronomy} {\bfseries 9} (Feb.,
  1966) 671}.

\bibitem{Bertotti:1966}
B.~{Bertotti}, \emph{{The Luminosity of Distant Galaxies}},
  \href{http://dx.doi.org/10.1098/rspa.1966.0203}{\emph{Proceedings of the
  Royal Society of London Series A} {\bfseries 294} (Sept., 1966) 195--207}.

\bibitem{Gunn:1967}
J.~E. {Gunn}, \emph{{On the Propagation of Light in Inhomogeneous Cosmologies.
  I. Mean Effects}}, \href{http://dx.doi.org/10.1086/149378}{\emph{Astrophys.
  J.} {\bfseries 150} (Dec., 1967) 737}.

\bibitem{Dyer:1974}
C.~C. {Dyer} and R.~C. {Roeder}, \emph{{Observations in Locally Inhomogeneous
  Cosmological Models}},
  \href{http://dx.doi.org/10.1086/152784}{\emph{Astrophys. J.} {\bfseries 189}
  (Apr., 1974) 167--176}.

\bibitem{Dyer:1981}
C.~C. {Dyer} and R.~C. {Roeder}, \emph{{On the transition from Weyl to Ricci
  focusing}}, \href{http://dx.doi.org/10.1007/BF00759864}{\emph{General
  Relativity and Gravitation} {\bfseries 13} (Dec., 1981) 1157--1160}.

\bibitem{Kasai:1990hd}
M.~Kasai, T.~Futamase and F.~Takahara, \emph{{Angular diameter distance in a
  clumpy universe}},
  \href{http://dx.doi.org/10.1016/0375-9601(90)90875-O}{\emph{Phys. Lett. A}
  {\bfseries 147} (1990) 97--105}.

\bibitem{Kibble:2004tm}
T.~W.~B. Kibble and R.~Lieu, \emph{{Average magnification effect of clumping of
  matter}}, \href{http://dx.doi.org/10.1086/444343}{\emph{Astrophys. J.}
  {\bfseries 632} (2005) 718--726},
  [\href{https://arxiv.org/abs/astro-ph/0412275}{{\ttfamily
  astro-ph/0412275}}].

\bibitem{Clifton:2009jw}
T.~Clifton and P.~G. Ferreira, \emph{{Archipelagian Cosmology: Dynamics and
  Observables in a Universe with Discretized Matter Content}},
  \href{http://dx.doi.org/10.1103/PhysRevD.84.109902}{\emph{Phys. Rev. D}
  {\bfseries 80} (2009) 103503},
  [\href{https://arxiv.org/abs/0907.4109}{{\ttfamily 0907.4109}}].

\bibitem{Clifton:2009bp}
T.~Clifton and P.~G. Ferreira, \emph{{Errors in Estimating $\Omega_\Lambda$ due
  to the Fluid Approximation}},
  \href{http://dx.doi.org/10.1088/1475-7516/2009/10/026}{\emph{JCAP} {\bfseries
  10} (2009) 026}, [\href{https://arxiv.org/abs/0908.4488}{{\ttfamily
  0908.4488}}].

\bibitem{Clifton:2010fr}
T.~Clifton, \emph{{Cosmology Without Averaging}},
  \href{http://dx.doi.org/10.1088/0264-9381/28/16/164011}{\emph{Class. Quant.
  Grav.} {\bfseries 28} (2011) 164011},
  [\href{https://arxiv.org/abs/1005.0788}{{\ttfamily 1005.0788}}].

\bibitem{Clarkson:2011br}
C.~Clarkson, G.~F.~R. Ellis, A.~Faltenbacher, R.~Maartens, O.~Umeh and J.-P.
  Uzan, \emph{{(Mis-)Interpreting supernovae observations in a lumpy
  universe}},
  \href{http://dx.doi.org/10.1111/j.1365-2966.2012.21750.x}{\emph{Mon. Not.
  Roy. Astron. Soc.} {\bfseries 426} (2012) 1121--1136},
  [\href{https://arxiv.org/abs/1109.2484}{{\ttfamily 1109.2484}}].

\bibitem{Clifton:2011mt}
T.~Clifton, P.~G. Ferreira and K.~O'Donnell, \emph{{An Improved Treatment of
  Optics in the Lindquist-Wheeler Models}},
  \href{http://dx.doi.org/10.1103/PhysRevD.85.023502}{\emph{Phys. Rev. D}
  {\bfseries 85} (2012) 023502},
  [\href{https://arxiv.org/abs/1110.3191}{{\ttfamily 1110.3191}}].

\bibitem{Bruneton:2012cg}
J.-P. Bruneton and J.~Larena, \emph{{Dynamics of a lattice Universe: The dust
  approximation in cosmology}},
  \href{http://dx.doi.org/10.1088/0264-9381/29/15/155001}{\emph{Class. Quant.
  Grav.} {\bfseries 29} (2012) 155001},
  [\href{https://arxiv.org/abs/1204.3433}{{\ttfamily 1204.3433}}].

\bibitem{Bruneton:2012ru}
J.-P. Bruneton and J.~Larena, \emph{{Observables in a lattice Universe}},
  \href{http://dx.doi.org/10.1088/0264-9381/30/2/025002}{\emph{Class. Quant.
  Grav.} {\bfseries 30} (2013) 025002},
  [\href{https://arxiv.org/abs/1208.1411}{{\ttfamily 1208.1411}}].

\bibitem{Larena:2012vn}
J.~Larena, \emph{{The fitting problem in a lattice Universe}},
  \href{http://dx.doi.org/10.1007/978-3-319-06761-2_53}{\emph{Springer Proc.
  Phys.} {\bfseries 157} (2014) 385--392},
  [\href{https://arxiv.org/abs/1210.2161}{{\ttfamily 1210.2161}}].

\bibitem{Liu:2015bya}
R.~G. Liu, \emph{{Lindquist-Wheeler formulation of lattice universes}},
  \href{http://dx.doi.org/10.1103/PhysRevD.92.063529}{\emph{Phys. Rev. D}
  {\bfseries 92} (2015) 063529},
  [\href{https://arxiv.org/abs/1501.05169}{{\ttfamily 1501.05169}}].

\bibitem{Sanghai:2015wia}
V.~A.~A. Sanghai and T.~Clifton, \emph{{Post-Newtonian Cosmological
  Modelling}}, \href{http://dx.doi.org/10.1103/PhysRevD.93.089903}{\emph{Phys.
  Rev. D} {\bfseries 91} (2015) 103532},
  [\href{https://arxiv.org/abs/1503.08747}{{\ttfamily 1503.08747}}].

\bibitem{Fleury:2015rwa}
P.~Fleury, J.~Larena and J.-P. Uzan, \emph{{The theory of stochastic
  cosmological lensing}},
  \href{http://dx.doi.org/10.1088/1475-7516/2015/11/022}{\emph{JCAP} {\bfseries
  11} (2015) 022}, [\href{https://arxiv.org/abs/1508.07903}{{\ttfamily
  1508.07903}}].

\bibitem{Bentivegna:2016fls}
E.~Bentivegna, M.~Korzy\'nski, I.~Hinder and D.~Gerlicher, \emph{{Light
  propagation through black-hole lattices}},
  \href{http://dx.doi.org/10.1088/1475-7516/2017/03/014}{\emph{JCAP} {\bfseries
  03} (2017) 014}, [\href{https://arxiv.org/abs/1611.09275}{{\ttfamily
  1611.09275}}].

\bibitem{Sanghai:2017yyn}
V.~A.~A. Sanghai, P.~Fleury and T.~Clifton, \emph{{Ray tracing and Hubble
  diagrams in post-Newtonian cosmology}},
  \href{http://dx.doi.org/10.1088/1475-7516/2017/07/028}{\emph{JCAP} {\bfseries
  07} (2017) 028}, [\href{https://arxiv.org/abs/1705.02328}{{\ttfamily
  1705.02328}}].

\bibitem{Bentivegna:2018koh}
E.~Bentivegna, T.~Clifton, J.~Durk, M.~Korzy\'nski and K.~Rosquist,
  \emph{{Black-Hole Lattices as Cosmological Models}},
  \href{http://dx.doi.org/10.1088/1361-6382/aac846}{\emph{Class. Quant. Grav.}
  {\bfseries 35} (2018) 175004},
  [\href{https://arxiv.org/abs/1801.01083}{{\ttfamily 1801.01083}}].

\bibitem{Fleury:2017owg}
P.~Fleury, J.~Larena and J.-P. Uzan, \emph{{Weak gravitational lensing of
  finite beams}},
  \href{http://dx.doi.org/10.1103/PhysRevLett.119.191101}{\emph{Phys. Rev.
  Lett.} {\bfseries 119} (2017) 191101},
  [\href{https://arxiv.org/abs/1706.09383}{{\ttfamily 1706.09383}}].

\bibitem{Fleury:2018cro}
P.~Fleury, J.~Larena and J.-P. Uzan, \emph{{Cosmic convergence and shear with
  extended sources}},
  \href{http://dx.doi.org/10.1103/PhysRevD.99.023525}{\emph{Phys. Rev. D}
  {\bfseries 99} (2019) 023525},
  [\href{https://arxiv.org/abs/1809.03919}{{\ttfamily 1809.03919}}].

\bibitem{Fleury:2018odh}
P.~Fleury, J.~Larena and J.-P. Uzan, \emph{{Weak lensing distortions beyond
  shear}}, \href{http://dx.doi.org/10.1103/PhysRevD.99.023526}{\emph{Phys. Rev.
  D} {\bfseries 99} (2019) 023526},
  [\href{https://arxiv.org/abs/1809.03924}{{\ttfamily 1809.03924}}].

\bibitem{Buchert:2011sx}
T.~Buchert and S.~R{\"a}s{\"a}nen, \emph{{Backreaction in late-time
  cosmology}},
  \href{http://dx.doi.org/10.1146/annurev.nucl.012809.104435}{\emph{Ann. Rev.
  Nucl. Part. Sci.} {\bfseries 62} (2012) 57--79},
  [\href{https://arxiv.org/abs/1112.5335}{{\ttfamily 1112.5335}}].

\bibitem{light_stat1}
S.~R{\"a}s{\"a}nen, \emph{{Light propagation in statistically homogeneous and
  isotropic dust universes}},
  \href{http://dx.doi.org/10.1088/1475-7516/2009/02/011}{\emph{JCAP} {\bfseries
  02} (2009) 011}, [\href{https://arxiv.org/abs/0812.2872}{{\ttfamily
  0812.2872}}].

\bibitem{light_stat2}
S.~R{\"a}s{\"a}nen, \emph{{Light propagation in statistically homogeneous and
  isotropic universes with general matter content}},
  \href{http://dx.doi.org/10.1088/1475-7516/2010/03/018}{\emph{JCAP} {\bfseries
  03} (2010) 018}, [\href{https://arxiv.org/abs/0912.3370}{{\ttfamily
  0912.3370}}].

\bibitem{Ellis:1971pg}
G.~F.~R. Ellis, \emph{{Relativistic cosmology}},
  \href{http://dx.doi.org/10.1007/s10714-009-0760-7}{\emph{Proc. Int. Sch.
  Phys. Fermi} {\bfseries 47} (1971) 104--182 (reprinted in Gen.Rel.Grav. 41
  (2009) 581--660)}.

\bibitem{Tsagas:2007yx}
C.~G. Tsagas, A.~Challinor and R.~Maartens, \emph{{Relativistic cosmology and
  large-scale structure}},
  \href{http://dx.doi.org/10.1016/j.physrep.2008.03.003}{\emph{Phys. Rept.}
  {\bfseries 465} (2008) 61--147},
  [\href{https://arxiv.org/abs/0705.4397}{{\ttfamily 0705.4397}}].

\bibitem{Lavinto:2013exa}
M.~Lavinto, S.~R{\"a}s{\"a}nen and S.~J. Szybka, \emph{{Average expansion rate
  and light propagation in a cosmological Tardis spacetime}},
  \href{http://dx.doi.org/10.1088/1475-7516/2013/12/051}{\emph{JCAP} {\bfseries
  12} (2013) 051}, [\href{https://arxiv.org/abs/1308.6731}{{\ttfamily
  1308.6731}}].

\bibitem{Bull:2012zx}
P.~Bull and T.~Clifton, \emph{{Local and non-local measures of acceleration in
  cosmology}}, \href{http://dx.doi.org/10.1103/PhysRevD.85.103512}{\emph{Phys.
  Rev. D} {\bfseries 85} (2012) 103512},
  [\href{https://arxiv.org/abs/1203.4479}{{\ttfamily 1203.4479}}].

\bibitem{Verde:2019ivm}
L.~Verde, T.~Treu and A.~G. Riess, \emph{{Tensions between the Early and the
  Late Universe}},
  \href{http://dx.doi.org/10.1038/s41550-019-0902-0}{\emph{Nature Astron.}
  {\bfseries 3} (7, 2019) 891},
  [\href{https://arxiv.org/abs/1907.10625}{{\ttfamily 1907.10625}}].

\bibitem{Aghanim:2018eyx}
{\scshape Planck} collaboration, N.~Aghanim et~al., \emph{{Planck 2018 results.
  VI. Cosmological parameters}},
  \href{https://arxiv.org/abs/1807.06209}{{\ttfamily 1807.06209}}.

\bibitem{Riess:2020fzl}
A.~G. Riess, S.~Casertano, W.~Yuan, J.~B. Bowers, L.~Macri, J.~C. Zinn et~al.,
  \emph{{Cosmic Distances Calibrated to 1\% Precision with Gaia EDR3 Parallaxes
  and Hubble Space Telescope Photometry of 75 Milky Way Cepheids Confirm
  Tension with $\Lambda$CDM}},
  \href{http://dx.doi.org/10.3847/2041-8213/abdbaf}{\emph{Astrophys. J. Lett.}
  {\bfseries 908} (2021) L6},
  [\href{https://arxiv.org/abs/2012.08534}{{\ttfamily 2012.08534}}].

\bibitem{Joos:1984uk}
E.~Joos and H.~D. Zeh, \emph{{The Emergence of classical properties through
  interaction with the environment}},
  \href{http://dx.doi.org/10.1007/BF01725541}{\emph{Z. Phys. B} {\bfseries 59}
  (1985) 223--243}.

\bibitem{Tegmark:1993yn}
M.~Tegmark, \emph{{Apparent wave function collapse caused by scattering}},
  \href{http://dx.doi.org/10.1007/BF00662807}{\emph{Found. Phys. Lett.}
  {\bfseries 6} (1993) 571},
  [\href{https://arxiv.org/abs/gr-qc/9310032}{{\ttfamily gr-qc/9310032}}].

\bibitem{Allali:2020ttz}
I.~Allali and M.~P. Hertzberg, \emph{{Gravitational Decoherence of Dark
  Matter}}, \href{http://dx.doi.org/10.1088/1475-7516/2020/07/056}{\emph{JCAP}
  {\bfseries 07} (2020) 056},
  [\href{https://arxiv.org/abs/2005.12287}{{\ttfamily 2005.12287}}].

\bibitem{Allali:2020shm}
I.~J. Allali and M.~P. Hertzberg, \emph{{Decoherence from General Relativity}},
  \href{http://dx.doi.org/10.1103/PhysRevD.103.104053}{\emph{Phys. Rev. D}
  {\bfseries 103} (2021) 104053},
  [\href{https://arxiv.org/abs/2012.12903}{{\ttfamily 2012.12903}}].

\bibitem{Allali:2021puy}
I.~J. Allali and M.~P. Hertzberg, \emph{{General Relativistic Decoherence with
  Applications to Dark Matter Detection}},
  \href{http://dx.doi.org/10.1103/PhysRevLett.127.031301}{\emph{Phys. Rev.
  Lett.} {\bfseries 127} (2021) 031301},
  [\href{https://arxiv.org/abs/2103.15892}{{\ttfamily 2103.15892}}].

\bibitem{RelCosmo}
G.~Ellis, R.~Maartens and M.~MacCallum, \emph{Relativistic Cosmology}.
\newblock Relativistic Cosmology. Cambridge University Press, 2012.

\end{thebibliography}\endgroup

\end{document}